\documentclass{article}
\usepackage[noadjust]{cite}
\usepackage{url}
\usepackage{verbatim}
\usepackage{graphicx}
\usepackage{tabularx}
\usepackage{lscape}
\usepackage{amsmath}
\usepackage[utf8]{inputenc}
\usepackage[lined,boxed, ruled]{algorithm2e}

\usepackage{amssymb}
\usepackage{pifont}
\begin{document}

\title{Dynamic Selection of Virtual Machines for Application Servers in Cloud Environments\footnote{
To cite this technical report, please use the following: Nikolay Grozev and Rajkumar Buyya, 
“Dynamic Selection of Virtual Machines for Application Servers in Cloud Environments,” 
Technical Report CLOUDS-TR-2016-1, Cloud Computing and Distributed Systems Laboratory, 
The University of Melbourne, February 7, 2016.}}

\author{
	\textbf{Nikolay Grozev}~and~\textbf{Rajkumar~Buyya}\\
	Cloud Computing and Distributed Systems (CLOUDS) Laboratory,\\
	Department of Computing and Information Systems,\\
	The University of Melbourne, Australia\\ [6mm]
	ngrozev@student.unimelb.edu.au,~rbuyya@unimelb.edu.au
}

\maketitle

\begin{abstract}
Autoscaling is a hallmark of cloud computing as it allows flexible just-in-time 
allocation and release of computational resources in response to dynamic and often 
unpredictable workloads. This is especially important for web applications whose workload 
is time dependent and prone to flash crowds. Most of them follow the 3-tier architectural 
pattern, and are divided into presentation, application/domain and data layers. In this 
work we focus on the application layer. Reactive autoscaling policies of the type 
\emph{``Instantiate a new Virtual Machine (VM) when the average server CPU utilisation 
reaches X\%''} have been used successfully since the dawn of cloud computing. But which 
VM type is the most suitable for the specific application at the moment remains an open 
question. In this work, we propose an approach for dynamic VM type selection. It uses a 
combination of online machine learning techniques, works in real time and adapts to 
changes in the users' workload patterns, application changes as well as middleware 
upgrades and reconfigurations. We have developed a prototype, which we tested with the 
CloudStone benchmark deployed on AWS EC2. Results show that our method quickly adapts to 
workload changes and reduces the total cost compared to the industry standard approach. 
\end{abstract}

\maketitle
\setlength{\textfloatsep}{5mm}

\section{Introduction}\label{sec:introduction}

Cloud computing is a disruptive IT model allowing enterprises to focus 
on their core business activities. Instead of investing in  their own IT 
infrastructures, they can now 
rent ready-to-use preconfigured virtual resources from cloud providers in a 
``pay-as-you-go'' manner. Organisations relying on fixed size private infrastructures 
often realise it can not match their dynamic needs, thus frequently being either under or 
overutilised. In contrast, in a cloud environment one can automatically acquire or 
release resources as they are needed --- a distinctive characteristic known as 
\emph{autoscaling}.   


This is especially important for large scale web applications, since the number of users 
fluctuates over time and is prone to flash crowds as a result of marketing campaigns and 
product releases. Most such applications follow the 3-tier architectural pattern and are 
divided in three standard layers/tiers~\cite{Fowler2003,Ramirez2000,Aarsten1996}:
\begin{itemize}
	\item \textbf{Presentation Layer} --- the end user interface.
	\item \textbf{Business/Domain Layer} --- implements the business logic. Hosted in one 
	or several Application Servers (AS).
	\item \textbf{Data Layer} --- manages the persistent data. Deployed in one or several 
	Database (DB) servers.
\end{itemize}

A user interacts with the presentation layer, which redirects the requests to an AS which 
in turn can access the data layer. The presentation layer is executed on the client's 
side (e.g. in a browser) and thus scalability is not an issue. Scaling the DB layer is a 
notorious challenge, since system architects have to balance between consistency, 
availability and partition tolerance following the results of the CAP 
theorem~\cite{Brewer2000,Brewer2012}. This field has already been well explored (Cattel 
surveys more than 20 related projects~\cite{Cattell2011}). Furthermore, Google has 
published about their new database which scales within and across data centres without 
violating transaction consistency~\cite{Corbett2013}. Hence data layer scaling is beyond 
the scope of our work. 

\begin{figure*}[!t]
	\centering
	
	\includegraphics[width=0.8 \linewidth]{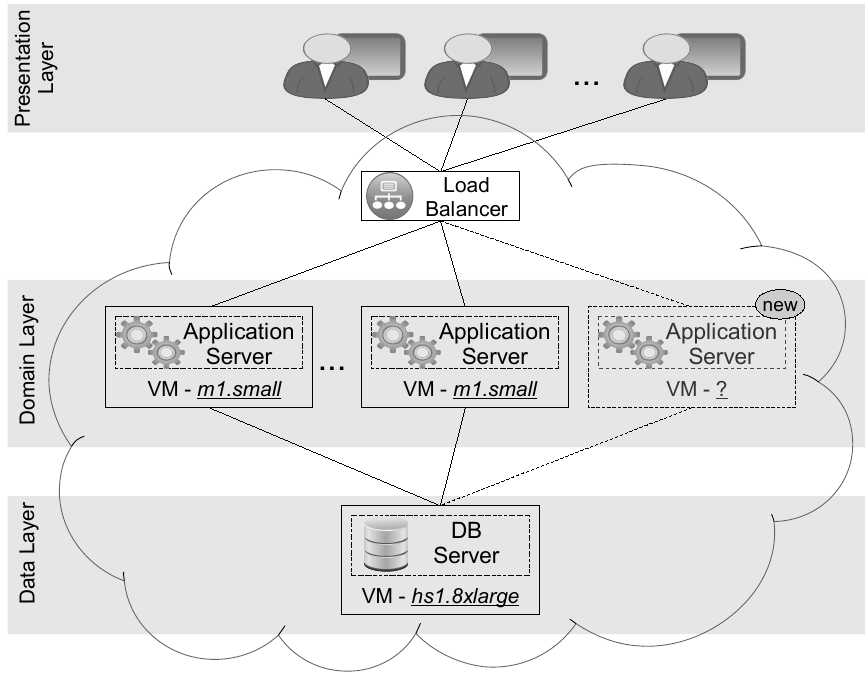} 
	\caption{A 3-tier application in Cloud. Whenever the autoscaling conditions are 
		activated, a new application server should be provisioned. In this work we select the 
		optimal VM type for the purpose.}
	\label{fig:3tier}
	
\end{figure*}

In general, autoscaling the Application Servers (AS) is comparatively straightforward. In 
an Infrastructure as a Service (IaaS) cloud environment, the AS VMs are deployed 
``behind'' a load balancer which redirects the incoming requests among them. Whenever the 
servers' capacity is insufficient, one or several new AS VMs are provisioned and 
associated with the load balancer and the DB layer --- see Figure~\ref{fig:3tier}.


\emph{But what should be the type of the new AS VM?} Most major cloud providers like 
Amazon EC2 and Google Compute Engine offer a predefined set of VM types with different 
performance capacities and prices. Currently, system engineers ``hardcode'' preselected 
VM types in the autoscaling rules based on their intuition or at best on historical 
performance observations. However, user workload characteristics vary over time leading 
to constantly evolving AS capacity requirements. For example, the proportion of browsing, 
bidding and buying requests in an e-commerce system can change significantly during a 
holiday season, which can change the server utilisation patterns. Middleware and 
operating system updates and reconfigurations can lead to changes in the utilisation 
patterns as well~\cite{Cherkasova2009}. This can also happen as a result of releasing new 
application features or updates.

Moreover, VM performance can vary significantly over time because of other VMs collocated 
on the same physical host causing resource contentions 
\cite{Tickoo2010,Dejun2009,Schad2010}. Hence even VM instances of the same type can 
perform very differently. From the viewpoint of the cloud's client this can not be 
predicted.

To illustrate better, let us consider a large scale web application with hundreds of 
dedicated AS VMs. Its engineers can analyse historical performance data to specify the 
most appropriate VM type in the autoscaling rules. However, they will have to reconsider 
their choice every time a new feature or a system upgrade is deployed. They will also 
have to constantly monitor for workload pattern changes and to react by adjusting the 
austoscaling rules. Given that VM performance capacities also vary over time, the job of 
selecting the most suitable VM type becomes practically unmanageable. This can result in 
significant financial losses, because of using suboptimal VMs.  


To address this, the key \textbf{contributions} of our work are (i) a machine learning 
approach which continuously learns the application's resource requirements and (ii) a 
dynamic VM type selection (DVTS) algorithm, which selects a VM type for new AS VMs. Since 
both workload specifics and VM performance vary over time, we propose an online approach, 
which learns the application's behaviour and the typical VM performance capacities in 
real time. It relieves system maintainers from having to manually reconfigure the 
autoscaling rules. 

The rest of the paper is organised as follows: In Section~\ref{related_work} we describe 
the related works. Section~\ref{overview} provides a succinct overview of our approach. 
Section~\ref{learning} discusses the machine learning approaches we employ to ``learn'' 
the application's requirements in real time. Section~\ref{selection} describes how to 
select an optimal VM type. Section~\ref{prototype} details the architecture of our 
prototype and the benchmark we use for evaluation. Section~\ref{experiments} describes 
our experiments and results. Finally, Section~\ref{conclusion} concludes and defines 
pathways for future work.

\section{Related Work} \label{related_work}


The area of static computing resource management has been well studied in the context 
of grids, clouds, and even multi-clouds~\cite{Tordsson2012}. However, the field of 
dynamic resource management in response to continuously varying workloads, 
which is especially important for web facing applications~\cite{Tordsson2012}, is still
in its infancy. Horizontal autoscaling policies are the predominant approach for dynamic
resource management and thus they have gained significant attention in recent years.

Lorido-Botran et al. classify autoscaling 
policies as \emph{reactive} and \emph{predictive} or 
\emph{proactive}~\cite{Lorido-Botran2012}. The most widely adopted \emph{reactive} 
approaches are based on threshold rules for performance metrics (e.g. CPU and RAM 
utilisation). For each such characteristic the system administrator provides a lower and 
upper threshold values. Resources are provisioned whenever an upper threshold is 
exceeded. Similarly, if a lower threshold is reached resources are released. How much 
resources are acquired or released when a threshold is reached is specified in user 
defined autoscaling rules. There are different ``flavours'' of threshold based 
approaches. For example in Amazon Auto Scaling~\cite{AmazonAutoScaling2013} one would 
typically use the average metrics from the virtual server farm, while 
RightScale~\cite{RightScaleWeb2012} provides a voting scheme, where thresholds are 
considered per VM and an autoscaling action is taken if the majority of the VMs ``agree'' 
on it. Combinations and extensions of both of these techniques have also been proposed 
\cite{Chieu2009,Chieu2011,Simmons2011}. \emph{Predictive} or \emph{proactive} approaches 
try to predict demand changes in order to allocate or deallocate resources. Multiple 
methods using approaches like reinforcement learning~\cite{Barrett2013,Dutreilh2011}, 
queuing theory~\cite{Ali-Eldin2012} and Kalman filters~\cite{Gandhi2014} to name a few 
have been proposed.

Our work is complementary to all these approaches. They indicate at what time resources 
should be provisioned, but do not select the resource type. Our approach selects the best 
resource (i.e. VM type) once it has been decided that the system should scale up 
horizontally.

Fernandez et al. propose a system for autoscaling web applications in 
clouds~\cite{Fernandez2014}. They monitor the performance of different VM types to 
infer their capacities. Our approach to this is different, as we inspect the available to 
each VM CPU capacity and measure the amount of ``stolen'' CPU instructions by the 
hypervisor from within the VM itself. This allows us to normalise the VMs' resource 
capacities to a common scale, which we use to compare them and for further analysis. 
Furthermore, their approach relies on a workload predictor, while ours is usable even in 
the case of purely reactive autoscaling.

Singh et al. use k-means clustering to analyse the workload mix (i.e. the different type 
of sessions) and then use a queueing model to determine each server's 
suitability~\cite{Singh2010}. However, they do not consider the performance 
variability of virtual machines, which we take into account. Also, they do not select the 
type of resource (e.g. VM) to provision and assume there is only one type, while this is 
precisely the focus of our work.

A part of our work is concerned with automated detection of application behaviour changes 
through a Hierarchical Temporal Memory (HTM) model. Similar work has been carried out by 
Cherkasova et al.~\cite{Cherkasova2009}, who propose a regression based anomaly 
detection approach. However, they analyse only the CPU utilisation. Moreover they 
consider that a set of user transactions' types is known beforehand. In contrast, our 
approach considers RAM as well and does not require application specific information like 
transaction types. Tan et al. propose the PREPARE performance anomaly detection 
system~\cite{Tan2012}. However, their approach can not be used by a cloud client, as 
it is built on top of the Xen virtual machine manager to which external clients have no 
access.   

Another part of our method is concerned with automatic selection of the \emph{learning 
rate} and \emph{momentum} of an artificial neural network (ANN). There is a significant 
amount of literature in this area as surveyed by Moreira and 
Fiesler~\cite{Moreira1995}. However, the works they overview are applicable for 
static data sets and have not been applied to learning from streaming online data whose 
patterns can vary over time. Moreover, they only consider how the intermediate parameters 
of the backpropagation algorithm vary and do not use additional domain specific logic. 
Although our approach is inspired by the work of Vogl et al.~\cite{Vogl1988} as it 
modifies the \emph{learning rate} and \emph{momentum} based on the prediction error, we 
go further and we modify them also based on the \emph{anomaly score} as reported by the 
Hierarchical Temporal Memory (HTM) models.

\section{Method Overview} \label{overview}

\begin{figure*}[!t]
	\centering
	
	\includegraphics[width=0.85 \linewidth]{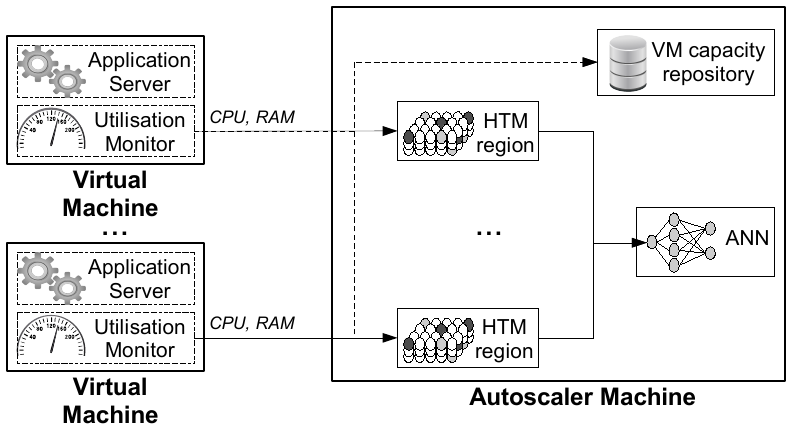} 
	\caption{System components and their interaction.}
	\label{fig:overview}
	
\end{figure*}

Figure~\ref{fig:overview} depicts an overview of our machine learning approach and how 
the system components interact. Within each AS VM we install a monitoring program which 
periodically records utilisation metrics. These measurements are transferred to an 
\emph{autoscaling component}, which can be hosted either in a cloud VM or on-premises. It 
is responsible for (i) monitoring AS VMs' performance (ii) updating machine learning 
models of the application behaviour and (iii) autoscaling.

Within each AS VM the \emph{utilisation monitors} report statistics about the CPU, RAM, 
disk and network card utilisations and the number of currently served users. These 
records are transferred every 5 seconds to the \emph{autoscaling} component, where they 
are normalised, as different VMs have different de facto resource capacities. In the 
machine learning approaches we only consider the CPU and RAM utilisations, as disk and 
network utilisations of AS VMs are typically small~\cite{Lloyd2013,Grozev2013}. 

For each AS VM the \emph{autoscaler} maintains a separate single-region Hierarchical 
Temporal Memory (HTM) model~\cite{Hawkins2011}, which is overviewed in a later section. 
In essence we use HTMs to detect changes in the application behaviour of each AS VM. We 
prefer HTM to other regression based anomaly detection approaches, as it can detect 
anomalies on a stream of multiple parameters (e.g. CPU and RAM). Whenever monitoring data 
is retrieved from an AS VM, the \emph{autoscaler} trains its HTM with the received number 
of users, CPU and RAM utilisations and outputs an \emph{anomaly score} defining how 
``unexpected'' the data is. 

As a next step we use these utilisation measurements to train a 3-tier artificial neural 
network (ANN) about the relationship between the number of served users and resource 
consumptions. We choose to use an ANN because of its suitability for online data streams. 
Other ``sliding window'' approaches operate only on a portion of the data stream. 
As a system's utilisation patterns can remain the same for long time intervals, the window
sizes may need to become impractically large or even be dynamically adjusted. 
On the contrary, an ANN does not operate on a 
fixed time window and is more adept with changes in the incoming data stream, as we will 
detail in a later section.

There is only one ANN and training samples from all AS VMs are used to train it. In 
essence the ANN represents a continuously updated regression model, which given a number 
of users predicts the needed resources to serve them within a single VM without causing 
resource contentions. Thus, we need to filter all training samples, which were taken 
during anomalous conditions (e.g. insufficient CPU or RAM capacity causing intensive 
context switching or disk swapping respectively). Such samples are not indicative of the 
relationship between number of users and the resource requirements in the absence of 
resource contentions. Furthermore, we use the \emph{anomaly score} of each training 
sample (extracted from HTM) to determine the respective \emph{learning speed} and 
\emph{momentum} parameters of the back propagation algorithm so that the ANN adapts 
quickly to changes in the utilisation patterns. 

Training the ANN and the HTMs happens online from the stream of VM measurements in 
parallel with the running application. Simultaneously we also maintain a \emph{VM 
capacity repository} of the latest VM capacity measurements. When a new VM is needed by 
the autoscaling component, we use this repository to infer the potential performance 
capacity of all VM types. At that time the ANN is already trained adequately and given 
the predicted performance capacities can be used to infer how many users each VM type 
could serve simultaneously. Based on that we select the VM type, with minimal cost to 
number of users ratio.

\section{Learning Application Behaviour} \label{learning}

\subsection{Utilisation Monitoring}

To measure VM performance utilisation, we use the \emph{SAR}, \emph{mpstat}, 
\emph{vmstat} and \emph{netstat} Linux monitoring tools. We use the mpstat~\emph{\%idle} 
metric to measure the percentage of time during which the CPU was idle. The 
\emph{\%steal} metric describes the percentage of ``stolen'' CPU cycles by a hypervisor 
(i.e. the proportion of time the CPU was not available to the VM) and can be used to 
evaluate the actual VM CPU capacity. Similarly, SAR provides the \emph{\%util} and 
\emph{\%ifutil} metrics as indicative of the disk's and network card's utilisations.

Measuring the RAM utilisation is more complex as operating systems keep in memory 
cached copies of recently accessed disk sectors in order to reduce disk access~\cite{Grozev2013}. 
Although in general this optimisation is essential for VM performance, web application 
servers (AS) are not usually I/O bound, as most of the application persistence is delegated
to the data base layer. Hence, using the \emph{vmstat} RAM utilisation metrics can 
be an overestimation of the actual memory consumption as it includes rarely accessed disk caches.
Thus, we use the \emph{``active memory''} \emph{vmstat} metric to measure memory consumption 
instead. It denotes the amount of recently used memory, which is unlikely to be claimed for other purposes.

Lastly, we need to evaluate the number of concurrently served users in an AS VM. This 
could be extracted from the AS middleware, but that would mean writing specific code for 
each type of middleware. Moreover, some proprietary solutions may not expose this 
information. Therefore, we use the number of distinct IP addresses with which the server 
has an active TCP socket, which can be obtained through the \emph{netstat} command. 
Typically, the AS VM is dedicated to running the AS and does not have other outgoing 
connections except for the connection to the persistence layer. Therefore, the number of 
addresses with active TCP sockets is a good measure of the number of currently served 
users. 


\subsection{Normalisation and Capacity Estimation}

Before proceeding to train the machine learning approaches, we need to normalise the 
measurements which  have different ``scales'', as the VMs have different RAM sizes and 
CPUs with different frequencies. Moreover, the actual CPU capacities within a single VM 
vary over time as a result of the dynamic collocation of other VMs on the same host. 

As a first step in normalising the CPU load, we need to evaluate the actual CPU capacity 
available to each VM. This can be extracted from the \emph{/proc/cpuinfo} Linux kernel 
file. If the VM has $n$ cores, \emph{/proc/cpuinfo} will list meta information about the 
physical CPU cores serving the VM including their frequencies $fr_1, ... fr_n$. The sum 
of these frequencies is the maximal processing capacity the VM can get, provided the 
hypervisor does not ``steal'' any processing time. Using the \emph{\%steal} mpstat 
parameter we can actually see what percentage of CPU operations have been taken away by 
the hypervisor. Subtracting this percentage from the sum of frequencies gives us the 
actual VM CPU capacity at the time of measurement. To normalise we further divide by the 
maximal CPU core frequency $fr_{max}$ multiplied by the maximal number of cores 
$n_{max\_cores}$ of all considered VMs in the cloud provider. This is a measure of the 
maximal VM CPU capacity one can obtain from the considered VM types. As clouds are made 
of commodity hardware, we will consider $fr_{max}=3.5GHZ$. This ensures that all values are
in the range $(0,1]$, although for some cloud providers all values may be much lower than 1,
depending on the underlying hardware they use. This is formalised in 
Eq.~\ref{eq:cpuCapacity}.

\begin{equation} \label{eq:cpuCapacity}
	cpuCapacityNorm = \frac{(100-\%steal) \sum\limits_{i=0}^{n} fr_i } { 100\ 
	n_{max\_cores}\ fr_{max} }
\end{equation}

Having computed the VM CPU capacity, we store it into the \emph{VM capacity repository}, 
so we can use it later on to infer the capacities of future VMs. Each repository record 
has the following fields:
\begin{itemize}
	\item \emph{time} - a time stamp of the capacity estimation;
	\item \emph{vm-type} - an identifier of the VM type - e.g. ``m1.small'';
	\item \emph{vm-id} - a unique identifier of the VM instance - e.g. its IP or elastic 
	DNS address;
	\item \emph{cpuCapacityNorm} - the computed CPU capacity.
\end{itemize}

If we further subtract the \emph{\%idle} percentage from the capacity we will get the 
actual CPU load given in  Eq.~\ref{eq:cpuNorm}.

\begin{equation} \label{eq:cpuNorm}
	cpuLoadNorm = \frac{(100-\%idle -\%steal) \sum\limits_{i=0}^{n} fr_i } { 100\ 
	n_{max\_cores}\ fr_{max} }
\end{equation}

Normalising the RAM load and capacity is easier, as they do not fluctuate like the CPU 
capacity. We divide the \emph{active memory} by the maximal amount of memory $RAM_{max}$ 
in all considered virtual machine types in the cloud - see Eq.~\ref{eq:ramNorm}.

\begin{equation} \label{eq:ramNorm}
	ramLoadNorm = \frac{active\textunderscore memory} {RAM_{max}}
\end{equation}

Whenever a new AS VM is needed, we have to estimate the CPU and RAM capacities of all 
available VM types based on the \emph{capacity repository} and their performance 
definitions provided by the provider. The normalised RAM capacity of a VM type is 
straightforward to estimate as we just need to divide the capacity in the provider's 
specification by $RAM_{max}$. To estimate the CPU capacity of a VM type we use the mean 
of the last 10 entries' capacities for this type in the \emph{capacity repository}. If 
there are no entries for this VM type in the repository (i.e. no VM of this type has been 
instantiated) we can heuristically extrapolate the CPU capacity from the capacities of 
the other VM types. Typically IaaS providers specify an estimation of each VM type's CPU 
capacity - e.g. Google Compute Engine Units (GCEU) in Google Compute Engine or Elastic 
Compute Units (ECU) in AWS. Hence given an unknown VM type $vmt$ we can extrapolate its 
normalised CPU capacity as:

\begin{multline} \label{eq:cpuCapEst}
cpuCapacity(vmt)=\\ \frac{1}{|V|} \sum\limits_{vmt_{i}\in V}^{ } \frac{cpuCapacity(vmt_{i}) \times cpuSpec(vmt_{i})} {cpuSpec(vmt)}
\end{multline}

Where $V$ is the set of VM types present in the \emph{capacity repository} and whose CPU 
capacity can be determined from previous measurements, $|V|$ is its cardinality, and 
$cpuSpec(vmt_{i})$ defines the 
cloud provider's estimation of a VM type's capacity - e.g. number of GCEUs or ECUs. 

\subsection{Anomaly Detection Through HTM}

\begin{figure}[t]
	\centering
	
	\includegraphics[width=0.75 \linewidth]{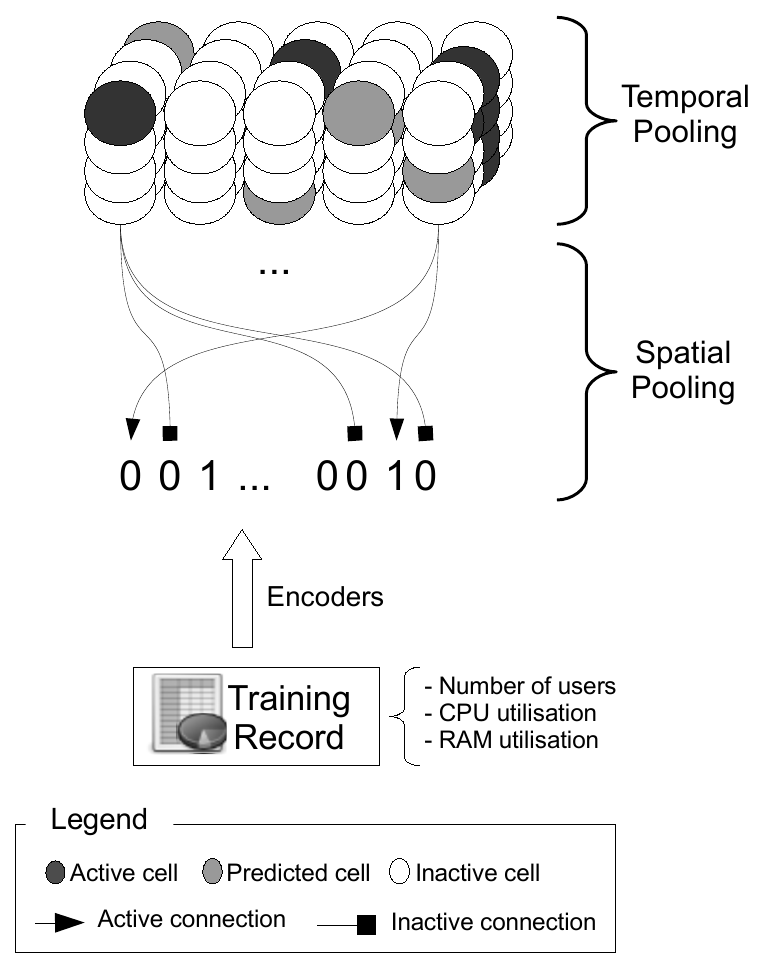} 
	\caption{HTM region structure.}
	\label{fig:htm}
	
\end{figure}

The Hierarchical Temporal Memory (HTM) model is inspired by the structure and 
organisation of the neocortex. It has been developed and commercialised by the Grok 
company~\cite{Grok2014} (formerly Numenta~\cite{Numenta2014}), and follows the 
concepts from Jeff Hawkins' book ``On Intelligence''~\cite{Hawkins2004}. The model creators 
build upon the seminal work of Mountcastle~\cite{Mountcastle1978} that the neocortex 
is predominantly uniform in structure and function even in regions handling different 
sensory inputs - e.g. visual, auditory, and touch. The HTM model tries to mimic this 
structure in a computational model. There are several differences compared to the 
biological structure of the neocortex in order to be computationally viable as described 
in the implementation white paper~\cite{Hawkins2011}.
Grok's implementation is available as an open source project called 
NuPIC~\cite{NuPIC2014}. In this section, we provide only a brief overview of HTM to 
introduce the reader to this concept. The interested reader is referred to the official 
documentation~\cite{Hawkins2011}.

HTMs consist of one or several stacked regions. During inference, input arrives into the 
lowest region, whose output serves as input to the successive one and so forth until the 
topmost region outputs the final result. The purpose of a region is to convert noisy 
input sequences to more stable abstract representations. Conceptually, the different 
regions represent different levels of abstraction in the learning process - i.e. the 
lowest level recognises low-level patterns, while each higher level layer recognises 
more complex ones based on the result of the previous one. 
In this work, we use single-region HTMs and we will focus on them in the rest of the 
section. 

A HTM region consists of columns of cells, which are most often arranged in a three 
dimensional grid - see Figure~\ref{fig:htm}. Each cell can be in one of three possible 
states: (i) active form feed forward input, (ii) active from lateral input (i.e. 
predicted), or (iii) inactive. Conceptually, active cells represent the state of the last 
input and predicted cells represent the likely state after future inputs. A HTM region 
receives as input a bit sequence. Special \emph{encoders} are used to convert input 
objects into bitwise representations, so that objects which are ``close'' in the sense of 
the target domain have similar bit representations. Upon receiving new binary input the 
HTM changes the states of the columns based on several rules summarised below.

As a first step the HTM has to decide which columns' cells will be activated for a given 
input - an algorithm known as \emph{Spatial Pooling}. It nullifies most of the 1 bits, so 
that only a small percentage (by default 2\%) are active. Each column is connected with a 
fixed sized (by default 50\% of the input length) random subset of input bits called the 
\emph{potential pool}. Each column's connection to an input bit has a ratio number in the 
range [0,1] associated with it known as the \emph{permanence}. HTM automatically adjusts 
the \emph{permanence} value of a connection after a new input record arrives, so that 
input positions whose value have been 0 or 1 and are members of the \emph{potential pool} 
of a selected column are decreased or increased respectively. Connections with 
\emph{permanences} above a predefined thresholds are considered active. Given an input, 
for each column the HTM defines its \emph{overlap score} as the number of active bits 
with active connections. Having computed this for every column, HTM selects a fixed sized 
(by default 2\%) set of columns with the highest \emph{overlap score}, so that no two 
columns within a predefined radius are active.

As a second step, HTM decides which cells within these columns to activate. This is 
called \emph{Temporal Pooling}. Within each of the selected columns the HTM activates 
only the cells which are in \emph{predicted} state. If there are no cells in predicted 
state within a column, then all of its cells are activated, which is also known as 
\emph{bursting}.

Next, the HTM makes a prediction of what its future state will be - i.e. which cells 
should be in predicted state. The main idea is that when a cell activates it establishes 
connections to the cells which were previously active. Each such connection is assigned a 
weight number. Over time if the two nodes of a connection become active in sequence 
again, this connection is strengthened, i.e. the weight is increased. Otherwise, the 
connection slowly decays, i.e. the weight is gradually decreased. Once a cell becomes 
active, all non-active cells having connections to it with weights above a certain 
threshold are assigned the predicted state. This is analogous to how synapses form and 
decay between neurons' dendrites in the neocortex in response to learning patterns.

The presence of predicted cell columns allows a HTM to predict what will be its likely 
state in terms of active cells after the next input. However, it also allows for the 
detection of anomalies. For example, if just a few predicted states become active this is 
a sign that the current input has not been expected. Thus the \emph{anomaly\_score} is 
defined as the proportion of active spatial pooler columns that were incorrectly 
predicted and is in the range $[0,1]$.

In our environment every 5 seconds we feed each HTM with a time stamp, the number of 
users and the CPU and RAM utilisations of the respective VM. We use the standard NuPIC 
scalar and date encoders to convert the input to binary input. As a result we get an 
\emph{anomaly score} denoting how expected the input is, in the light of the previously 
described algorithms. 

\subsection{ANN Training}

\begin{figure}[!t]
	\centering
	
	\includegraphics[width=0.75 \linewidth]{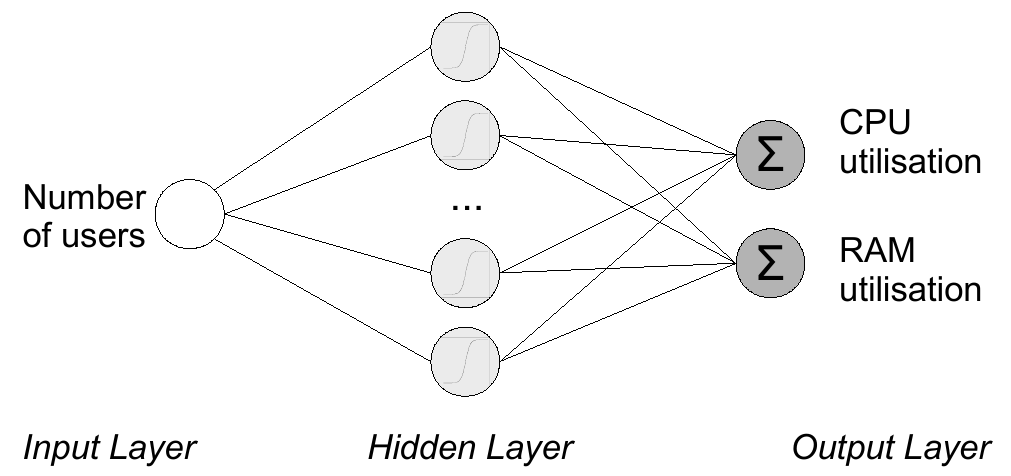} 
	\caption{ANN topology.}
	\label{fig:ANN}
	
\end{figure}

Figure~\ref{fig:ANN} depicts the topology of the artificial neural network (ANN). It has 
one input --- the number of users. The hidden layer has 250 neurons with the sigmoid 
activation function. The output layer has two output nodes with linear activation 
functions, which predict the normalised CPU and RAM utilisations within an AS VM.

Once a VM's measurements are received and normalised and the \emph{anomaly score} is 
computed by the respective HTM region, the ANN can be trained. As discussed, we need to 
filter out the VM measurements which are not representative of normal, contention free 
application execution, in order to ``learn'' the ``right'' relationship between number of 
users and resource utilisations. We filter all VM measurements in which the CPU, RAM, 
hard disk or network card utilisations are above a certain threshold (e.g. 70\%). 
Similarly, we filter measurements with negligible load --- i.e. less than 25 users or less 
than 10\% CPU utilisation. We also ignore measurements from periods during which the 
number of users has changed significantly --- e.g. in the beginning of the period there 
were 100 users and at the end there were 200. Such performance observations are not 
indicative of an actual relationship between number of users and resource utilisations. 
Thus, we ignore measurements for which the number of users is less than 50\% or more than 
150\% of the average of the previous 3 measured numbers of users from the same VM.

Since we are training the ANN with streaming data, we need to make sure it is not 
overfitted to the latest training samples. For example if we have constant workload for a 
few hours we will be receiving very similar training samples in the ANN during this 
period. Hence the ANN can become overfitted for such samples and lose its fitness for the 
previous ones. To avoid this problem, we filter out measurements/training samples, which 
are already well predicted. More specifically, if a VM measurement is already predicted 
with a \emph{root mean square error} (RMSE) less than $0.01$ it is filtered out and the 
ANN is not trained with it. We call this value $rmse^{pre}$ because it is obtained for 
each training sample before the ANN is trained with it. It is computed as per 
Eq.~\ref{eq:RMSE}, where $output_{i}$ and $expected_{i}$ are the values of the output 
neurons and the expected values respectively.

\begin{equation} \label{eq:RMSE}
	rmse^{pre} = \sqrt {\sum (output_{i} - expected_{i}) ^{2}}
\end{equation}

With each measurement, which is not filtered out, we perform one or several 
iterations/epochs of the back-propagation algorithm with the number of users as input and 
the normalised CPU and RAM utilisations as expected output. The back-propagation 
algorithm has two important parameters --- the \emph{learning rate} and the 
\emph{momentum}. In essence, the \emph{learning rate} is a ratio number in the interval 
$(0,1)$ which defines the amount of weight update in the direction of the gradient 
descent for each training sample~\cite{Moreira1995}. For each weight update, the 
\emph{momentum} term defines what proportion of the previous weight update should be 
added to it. It is also a ratio number in the interval $(0,1)$. Using a \emph{momentum} 
the neural network becomes more resilient to oscillations in the training data by 
``damping'' the optimisation procedure~\cite{Moreira1995}.

For our training environment we need a low \emph{learning rate} and a high 
\emph{momentum}, as there are a lot of oscillations in the incoming VM measurements. We 
select the \emph{learning rate} to be $lr=0.001$ and the \emph{momentum} $m=0.9$. We call 
these values the \emph{ideal parameters}, as these are the values we would like to use 
once the ANN is close to convergence. However, the low \emph{learning rate} and high 
\emph{momentum} result in slow convergence in the initial stages, meaning that the ANN 
may not be well trained before it is used. Furthermore, if the workload pattern changes, 
the ANN may need a large number of training samples and thus time until it is tuned 
appropriately. Hence the actual \emph{learning rate} and \emph{momentum} must be defined 
dynamically.

One approach to resolve this is to start with a high \emph{learning rate} and low 
\emph{momentum} and then respectively decrease/increase them to the desired 
values~\cite{Moreira1995,Vogl1988}. This allows the back-propagation algorithm to 
converge more rapidly during the initial steps of the training. We define these 
parameters in the initial stages using the asymptotic properties of the sigmoid function, 
given in Eq.~\ref{eq:sigmoid}.

\begin{equation} \label{eq:sigmoid}
	s(x) = \frac{1}{1 - e^{-x}} 
\end{equation} 

As we need to start with a high \emph{learning rate} and then decrease it gradually to 
$lr$, we could define the learning rate $lr_{k}$ for the $k$-th training sample as 
$s(-k)$. However, the sigmoid function decreases too steeply for negative integer 
parameters and as a result the learning rate is higher than $lr$ for just a few training 
samples. To solve this we use the square root of $k$ instead and thus our first 
approximation of the \emph{learning rate} is:

\begin{equation} \label{eq:lr1k}
	lr^{(1)}_{k} = max(lr, s(-\sqrt{k})) 
\end{equation} 

\begin{figure}[!t]
	\centering
	
	\includegraphics[width=0.85 \linewidth]{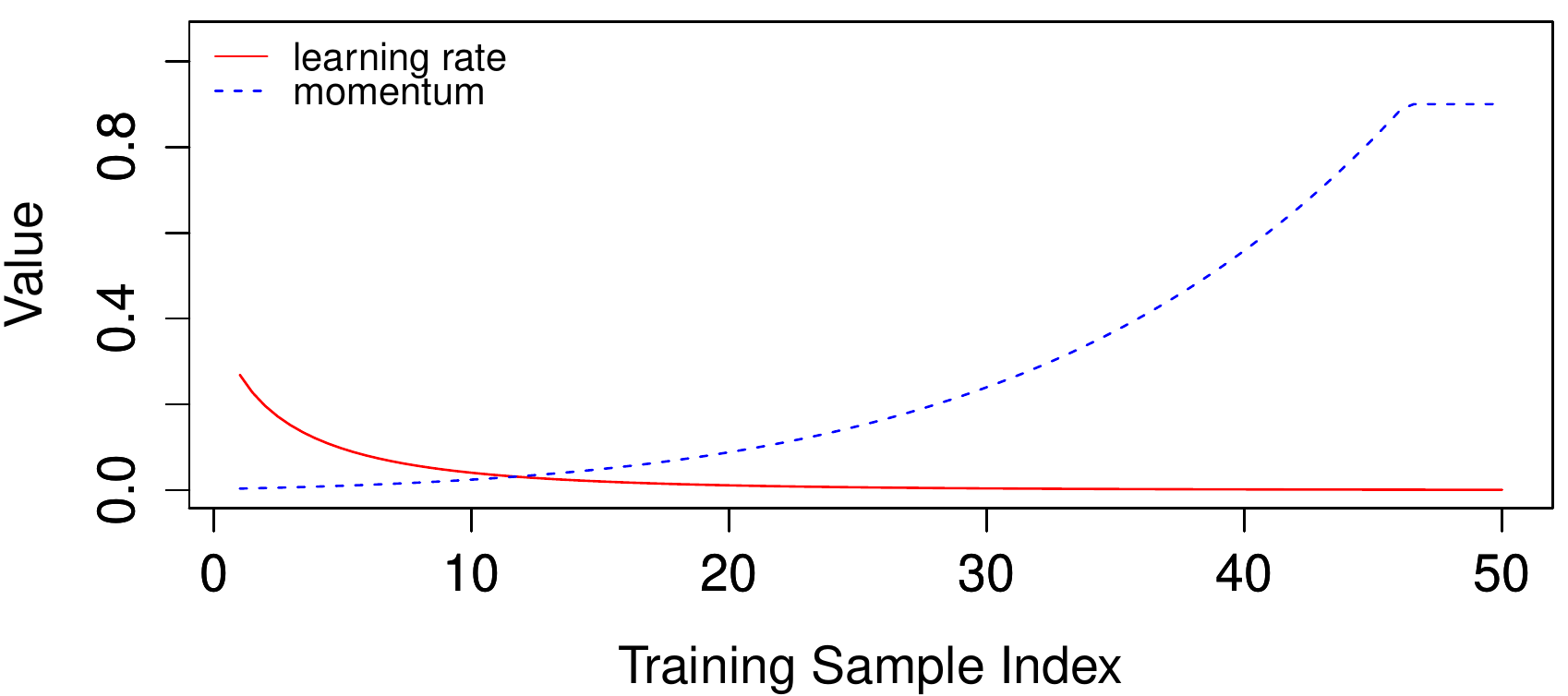} 
	\caption{The $lr^{(1)}_{k}$ approximation of the \emph{learning rate} and the 
	respective momentum during the initial ANN training stages.}
	\label{fig:LR-Mom}
	
\end{figure}

As a result $lr^{(1)}_{k}$ gradually decreases as more training samples arrive. 
Figure~\ref{fig:LR-Mom} depicts how it changes over time. 

We also need to ensure that it increases in case unusual training data signalling a 
workload change arrives and thus we need to elaborate $lr^{(1)}_{k}$. For this we keep a 
record of the last 10 samples' \emph{anomaly scores} and errors (i.e. $rmse^{pre}$). The 
higher the latest anomaly scores, the more ``unexpected'' the samples are and therefore 
the \emph{learning rate} must be increased. Similarly, the higher the sample's 
$rmse^{pre}$ compared to the previous errors, the less fit for it the ANN is and thus the 
\emph{learning rate} must be increased as well. Thus our second elaborated approximation 
of the \emph{learning rate} is:

\begin{equation} \label{eq:lr2k}
	lr^{(2)}_{k} = lr^{(1)}_{k} \ max(1, \frac{rmse^{pre}_{k}}{\overline{rmse}}) 
	\prod_{i=0}^{9} 2s(an_{k-i})
\end{equation}

where $an_{k}$ and $rmse^{pre}_{k}$ are the \emph{anomaly score} and the error of the 
$k$-th sample and $\overline{rmse}$ is the average error of the last 10 samples. Note 
that we use the sigmoid function for the anomaly scores in order to diminish the effect 
of low values.

In some cases the \emph{learning rate} can become too big in the initial training 
iterations, which will in fact hamper the convergence. To overcome this problem, for each 
sample $k$ we run a training iteration with $lr^{(2)}_{k}$, compute its RMSE 
$rmse^{post}_{k}$ and then revert the results of this iteration. By comparing 
$rmse^{pre}_{k}$ and $rmse^{post}_{k}$ we can see if training with this $lr^{(2)}_{k}$ 
will contribute to the convergence~\cite{Vogl1988}. If not, we use the ideal parameter 
$lr$ instead. Thus we finally define the \emph{learning rate} parameter $lr_{k}$ in 
Eq.~\ref{eq:lrk}:

\begin{equation} \label{eq:lrk}
	lr_{k} =
	\left\{
	\begin{array}{ll}
		lr^{(2)}_{k}  & \mbox{if}\ rmse^{pre}_{k} > rmse^{post}_{k} \\
		lr		 & \mbox{otherwise}
	\end{array}
	\right.
\end{equation}
%
%
%

Similarly we have to gradually increase the \emph{momentum} as we decrease the 
\emph{learning rate} until the ideal \emph{momentum} is reached. If a workload change is 
present we need to decrease the \emph{momentum} in order to increase the learning speed. 
Hence, we can just use the ratio of the ideal learning rate $lr$ to the current one as 
shown in Eq.~\ref{eq:mk}. 

\begin{equation} \label{eq:mk}
	m_{k} = min(m, \frac{lr}{lr^{(2)}_{k}}) 
\end{equation}

Figure~\ref{fig:LR-Mom} depicts how the \emph{learning rate} and \emph{momentum} change 
during the initial training stages, given there are no anomalies, accuracy losses and 
$\forall k:rmse^{pre}_{k} > rmse^{post}_{k} $ --- i.e. when $\forall k: lr^{(1)}_{k} = 
lr^{(2)}_{k} = lr_{k}$. Figure~\ref{fig:LR-Init} shows the actual $lr_{k}$ given 
realistic workload.

Furthermore, to speed up convergence it is beneficial to run multiple \emph{epochs} (i.e. 
repeated training iterations) with the first incoming samples and with samples taken 
after a workload change. The ideal \emph{learning rate} $lr$ and its approximation 
$lr^{(2)}_{k}$ already embody this information and we could simply use their ratio. 
However, $\frac{lr^{(2)}_{k}}{lr}$ can easily exceed 300 given $lr=0.001$, resulting in 
over-training with particular samples. Hence we take the logarithm of it as in 
Eq.~\ref{eq:ek}:

\begin{equation} \label{eq:ek}
	e_{k} = \left \lfloor{ 1 + ln(\frac{lr^{(2)}_{k}}{lr}) }\right \rfloor 
\end{equation}

\section{Virtual Machine Type Selection} \label{selection}

\begin{algorithm}[!t]
	
	\LinesNumbered
	\SetKwInOut{Input}{input}\SetKwInOut{Output}{output}
	
	\SetCommentSty{textnormal}
	
	\Input{$VT$, $ann$, $\Delta$, $minU$, $maxU$}
	
	\BlankLine
	
	$bestVmt \longleftarrow $ null\;
	$bestCost \longleftarrow $ 0\;
	
	\BlankLine
	\For{$vmt \in VT$ \tcp*{Inspect all VM types}} {
		$cpuCapacity \longleftarrow $ $vmt$'s norm. CPU capacity \;
		$ramCapacity \longleftarrow $ $vmt$'s norm. RAM capacity\;
		$vmtCost \longleftarrow $ $vmt$'s cost per time unit\;
		\BlankLine
		$userCapacity \longleftarrow $ 0\;
		$n \longleftarrow minU$\;
		
		\While{True \tcp*{Find how many users it can take}} { 
			$cpu,ram \longleftarrow predict(ann, n, minU, maxU)$\;
			\uIf{$cpu < cpuCapacity$ and $ram < ramCapacity$} {
				$userCapacity \longleftarrow n$\;
			} \Else  {
			break\;
		}
		$n \longleftarrow n+\Delta$\;
	}
	\BlankLine
	\tcp{Approximate the cost for a user per time unit}	
	$userCost \longleftarrow $ $\frac{vmtCost}{userCapacity} $\;
	
	\BlankLine
	\tcp{Find the cheapest VM type}	
	\If{$userCost < bestCost$} {
		$bestCost \longleftarrow userCost$\;
		$bestVmt \longleftarrow vmt$\;
	}
}
return $bestVmt$\;

\caption{ Dynamic VM Type Selection (DVTS)}
\label{algo:vmtselect}
\end{algorithm}


When a new VM has to be provisioned the ANN should be already trained so that we can 
estimate the relationship between number of users and CPU and RAM requirements. The 
procedure is formalised in Algorithm~\ref{algo:vmtselect}. We loop over all VM types~$VT$ 
(line 3) and for each one we estimate its normalised CPU and RAM capacity based on the 
\emph{capacity repository} as explained earlier (lines 5-6). The VM cost per time unit 
(e.g. hour in AWS or minute in Google Compute Engine) is obtained from the provider's 
specification (line~7).

Next we approximate the number of users that a VM of this type is expected to be able to 
serve (lines 10-18). We iteratively increase $n$ by $\Delta$ starting from $minU$, which 
is the minimal number of users we have encountered while training the neural network. We 
use the procedure $predict$ (defined separately in Algorithm~\ref{algo:utilest}) to 
estimate the normalised CPU and RAM demands that each of these values of $n$ would cause. 
We do so until the CPU or RAM demands exceed the capacity of the inspected VM type. Hence, 
we use the previous value of $n$ as an estimation of the number of users a VM of that 
type can accommodate. Finally, we select the VM type with the lowest cost to number of 
users ratio (lines 20-23).

Algorithm~\ref{algo:utilest} describes how to predict the normalised utilisations caused 
by $n$ concurrent users. If $n$ is less than the maximum number of users $maxU$ we 
trained the ANN with, then we can just use the ANN's prediction (line 5). However, if $n$ 
is greater than $maxU$ the ANN may not predict accurately. For example if we have used a 
single \emph{small} VM to train the ANN, and then we try to predict the capacity of a 
\emph{large} VM, $n$ can become much larger than the entries of the training data and the 
regression model may be inaccurate. Thus, we extrapolate the CPU and RAM requirements 
(lines 7-11) based on the range of values we trained the ANN with and the performance 
model we have proposed in a previous work~\cite{Grozev2013}.

\begin{algorithm}[!t]
	
	\LinesNumbered
	\SetKwInOut{Input}{input}\SetKwInOut{Output}{output}
	
	\SetCommentSty{textnormal}
	
	\Input{$ann$, $n$, $minU$, $maxU$}
	
	\BlankLine
	$cpu \longleftarrow $ 0\;
	$ram \longleftarrow $ 0\;
	
	\BlankLine
	\uIf{$n < maxUsers$ \tcp*{If within range - use ANN}} {
		$cpu, ram \longleftarrow $ $ann.run(n)$\;
	} \Else   { 
	\tcp{If outside range - extrapolate}
	\BlankLine
	$minRam, minCPU \longleftarrow $ $ann.run(minU)$\;
	$maxRam, maxCPU \longleftarrow $ $ann.run(maxU)$\;
	$cpuPerUser \longleftarrow $ $\frac{(maxCPU-minCPU)}{(maxU - minU)}$\;	
	$ramPerUser \longleftarrow $ $\frac{(maxRam-minRam)}{(maxU - minU)}$\;
	$cpu \longleftarrow maxCPU + cpuPerUser (n-maxU)$
	$ram \longleftarrow maxCPU + ramPerUser (n-maxU)$
}
return $cpu, ram$\;

\caption{Resource Utilisation Estimation}
\label{algo:utilest}
\end{algorithm}


\section{Benchmark and Prototype} \label{prototype}

There are two main approaches for experimental validation of a distributed system's 
performance --- through a simulation or a prototype. Discrete event simulators like 
CloudSim~\cite{Calheiros2011} have been used throughout industry and academia to quickly 
evaluate scheduling and provisioning approaches for large scale cloud infrastructure 
without having to pay for expensive test beds. Unfortunately, such simulators work on a 
simplified cloud performance model and do not represent realistic VM performance 
variability, which is essential for testing our system. Moreover, simulations can be 
quite inaccurate when the simulated system serves resource demanding workloads, as they 
do not consider aspects like CPU caching, disk data caching in RAM and garbage 
collection~\cite{Grozev2013}. Therefore, we test our method through a prototype and a 
standard benchmark deployed in a public cloud environment.

\begin{figure*}[!t]
	\centering
	
	\includegraphics[width=1.05 \linewidth]{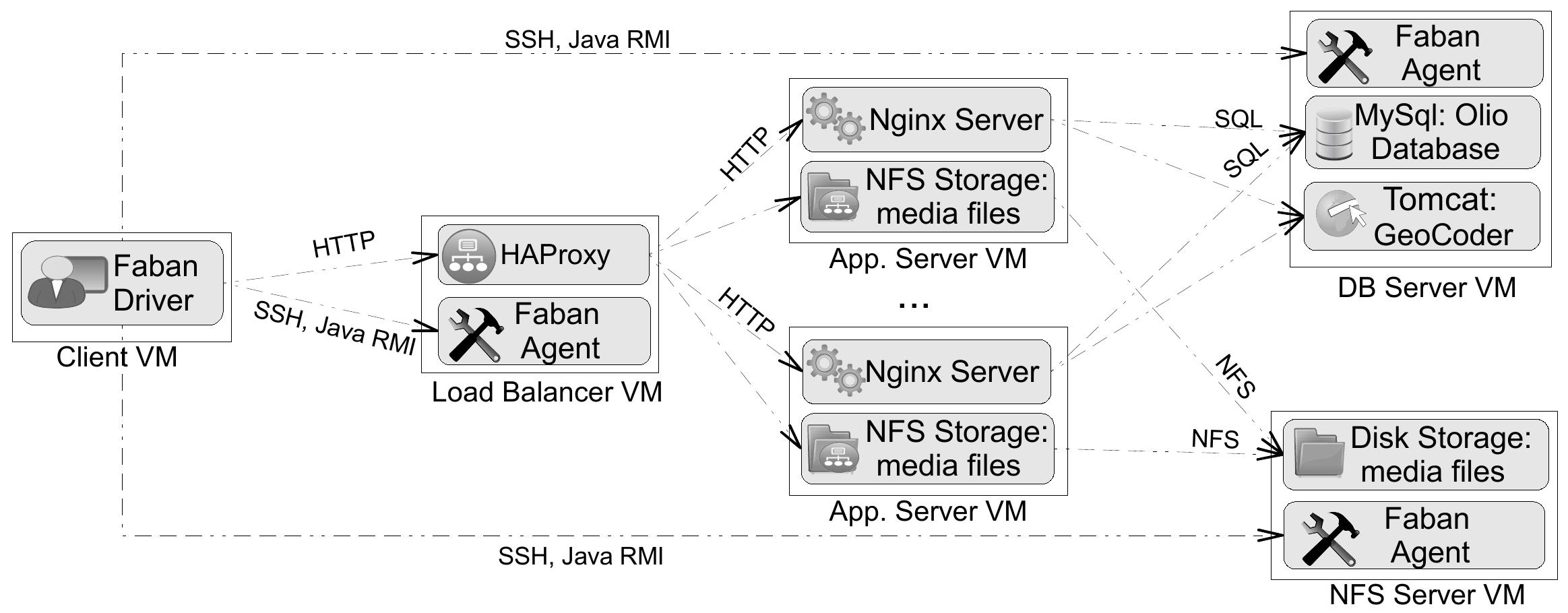} 
	\caption{CloudStone benchmark's extended topology.}
	\label{fig:ClouStone}
	
\end{figure*}

We validate our approach with the CloudStone~\cite{CloudSuite2014,Sobel2008} web 
benchmark deployed in Amazon AWS. It follows the standard 3-tier architecture. By default 
CloudStone is not scalable, meaning that it can only use a single AS. Thus we had to 
extend it to accommodate multiple servers. Our installation scripts and configurations 
are available as open source code. For space considerations we will not discuss these 
technical details and will only provide an overview. The interested readers can refer to 
our online documentation and installation 
instructions.\footnote{http://nikolaygrozev.wordpress.com/2014/06/02/advanced-automated-cloudstone-setup-in-ubuntu-vms-part-2/}

The benchmark deployment topology is depicted in Figure~\ref{fig:ClouStone}. CloudStone 
uses the \emph{Faban} harness to manage the runs and to emulate users. The \emph{faban 
driver}, which is deployed in the client VM communicates with the \emph{faban agents} 
deployed in other VMs to start or stop tests. It also emulates the incoming user requests 
to the application. These requests arrive at a HAProxy \emph{load balancer} which 
distributes them across one or many application servers (AS). CloudStone is based on the 
Olio application, which is a PHP social network website deployed in a Nginx server. In 
the beginning we start with a single AS ``behind'' the \emph{load balancer}. When a new 
AS VM is provisioned we associate it with the \emph{load balancer}. We update its 
weighted round robin policy, so that incoming request are distributed among the AS VMs 
proportionally to their declared CPU capacity (i.e. ECU).

The persistent layer is hosted in a MySql server deployed within a separate DB VM. 
CloudStone has two additional components - (i) a geocoding service called 
\emph{GeoCoder}, hosted in an Apache Tomcat server and (ii) a shared \emph{file storage} 
hosting media files. They are both required by all application servers. We have deployed 
the geocoding service in the DB VM. The file storage is deployed in a Network File System 
(NFS) server on a separate VM with 1TB EBS storage, which is mounted from each AS VM.

We use ``m3.medium'' VMs for the client, load balancer and DB server and ``m1.small'' for 
the NFS server. The types of the AS VMs are defined differently for each experiment. All 
VMs run 64bit Ubuntu Linux 14.04.

Our prototype of an autoscaling component is hosted on an on-premises physical machine 
and implements the previously discussed algorithms and approaches. It uses the 
JClouds~\cite{JCloudsWeb2012} multi-cloud library to provision resources, and thus 
can be used in other clouds as well. We use the NuPIC~\cite{NuPIC2014} and 
FANN~\cite{FANN2014} libraries to implement HTM and ANN respectively. We ignore the 
first 110 \emph{anomaly scores} reported from the HTM, as we observed that these results 
are inaccurate (i.e. always 1 or 0) until it receives initial training. Whenever a new AS 
VM is provisioned we initialise it with a deep copy of the HTM of the first AS VM, which 
is the most trained one. The monitoring programs deployed within each VM are implemented 
as bash scripts, and are accessed by the autoscaling component through SSH. Our 
implementation of Algorithm~\ref{algo:utilest} uses $\Delta=5$. 

Previously we discussed that the number of current users could be approximated by 
counting the number of distinct IP addresses to which there is an active TCP session. 
However, in CloudStone all users are emulated from the same client VM and thus have the 
same source IP address. Thus, we use the number of recently modified web server session 
files instead.

Our autoscaling component implementation follows the Amazon Auto 
Scaling~\cite{AmazonAutoScaling2013} approach and provisions a new AS VM once the 
average utilisation of the server farm reaches 70\% for more than 10 seconds. 
Hence, we ensure that in all experiments the AS VMs are not overloaded. Thus, even if 
there are SLA violations, they are caused either by the network or the DB layer, and the 
AS layer does not contribute to them. We also implement a \emph{cool down} period of 10 
minutes. 

\section{Validation} \label{experiments}

In our experiments, we consider three VM types: \emph{m1.small}, \emph{m1.medium} and 
\emph{m3.medium}. Table~\ref{tbl:vmt} summarises their cost and declared capacities in 
the Sydney AWS region which we use.


\begin{table}[!t]
	\centering
	
	\caption{AWS VM type definitions.}
		\begin{tabular}{l|rll}
			
			\textbf{VM type} & \multicolumn{1}{l}{\textbf{ECU}} & \textbf{RAM} & 
			\textbf{Cost per hour} \\ 
			\hline
			m1.small & 1 & 1.7GB & \$0.058 \\ 
			m1.medium & 2 & 3.75GB & \$0.117 \\ 
			m3.medium & 3 & 3.75GB & \$0.098 \\  
		\end{tabular}

	\label{tbl:vmt}
	
\end{table}


In all experiments we use the same workload. We start by emulating 30 users and each 6 
minutes we increase the total number of users with 10 until 400 users are reached. To 
achieve this we run a sequence of CloudStone benchmarks, each having 1 minute ramp-up and 
5 minutes steady state execution time. Given CloudStone's start-up and shut-down times, 
this amounts to more than 5 hours per experiment. The goal is to gradually increase the 
number of users, thus causing the system to scale up multiple times. 

To test our approach in the case of a workload characteristic change we ``inject'' such a 
change 3.5 hours after each experiment's start. To do so we manipulate the 
\emph{utilisation monitors} to report higher values. More specifically they increase the 
reported CPU utilisations with 10\% and the reported RAM utilisation with 1GB plus 2MB 
for every currently served user.  

We implement one experiment, which is initialised with a \emph{m1.small} AS VM and each 
new VM's type is chosen based on our method (DVTS). We also execute 3 baseline 
experiments, each of which statically selects the same VM type whenever a new VM is 
needed, analogously to the standard AWS Auto Scaling rules. 

\begin{figure}[!t]
	\centering
	
	\includegraphics[width=0.8 \linewidth]{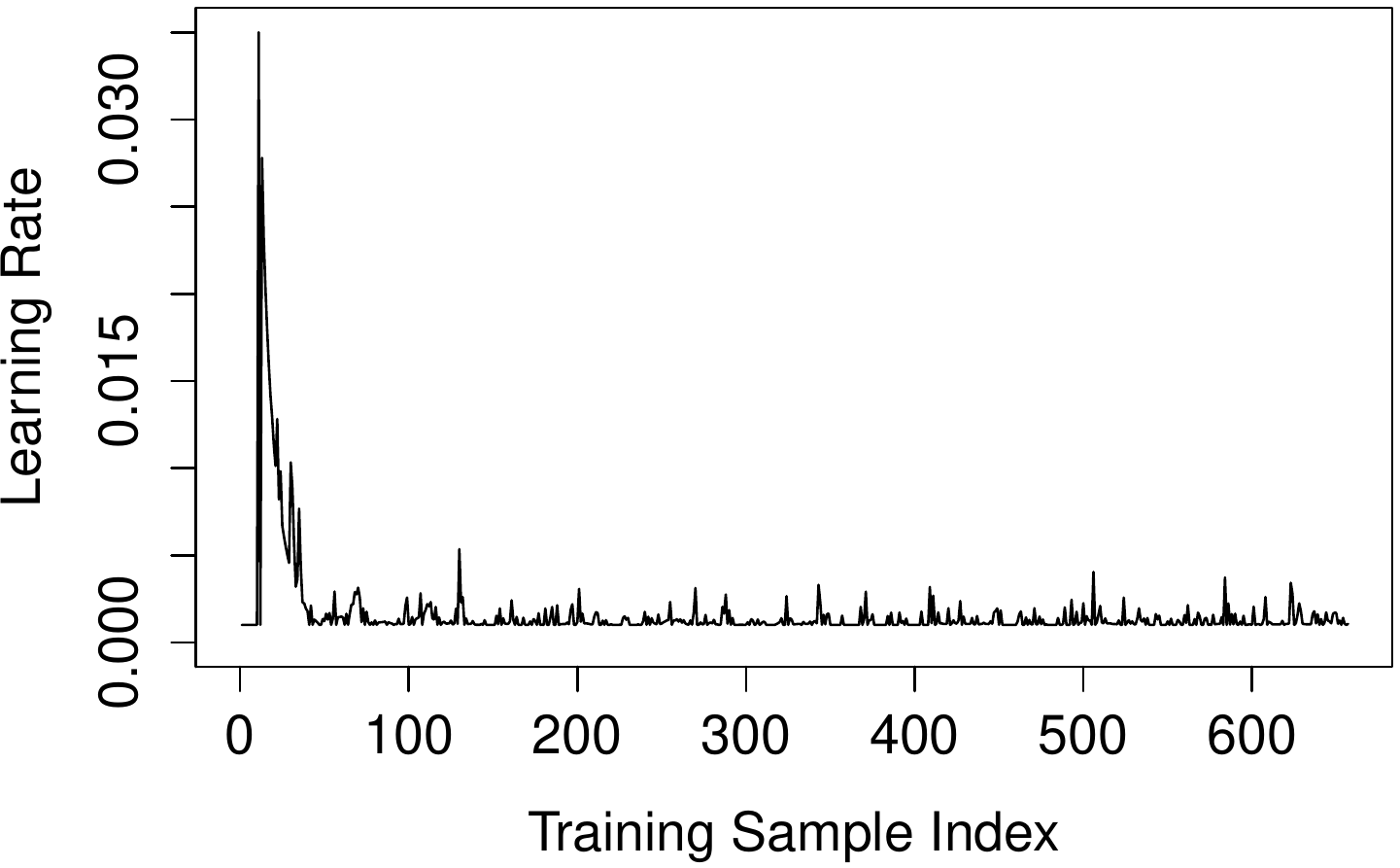} 
	\caption{Learning rate $lr_{k}$ during initial stages of training the ANN.}
	\label{fig:LR-Init}
	
\end{figure}

First we investigate the behaviour of DVTS before the workload change. It continuously 
trains one HTM for the first AS VM and the ANN. In the initial stages the ANN 
\emph{learning rate} and \emph{momentum} decrease and increase respectively to facilitate 
faster training. For example, the \emph{learning rate} $lr_{k}$ (defined in 
Eq.~\ref{eq:lrk}) during the initial stages is depicted in Fig~\ref{fig:LR-Init}. It 
shows how $lr_{k}$ drastically reduces as the ANN improves its accuracy after only a few 
tens of training samples. Once the AS VM gets overloaded we select a new VM type. At this 
point we only have information about \emph{m1.small} in the \emph{capacity repository} 
and therefore we infer the other CPU capacities based on Eq.~\ref{eq:cpuCapEst}. Finally 
using Algorithm~\ref{algo:vmtselect} we select \emph{m3.medium} as the type for the 
second VM. 

After the new VM is instantiated, the autoscaling component starts its monitoring. It 
trains the ANN and a new dedicated HTM with its measurements. It also updates the 
\emph{capacity repository} with the CPU capacity of the new VM. Surprisingly, we observe 
that on average its CPU capacity is about 35\% better than the one of the \emph{m1.small} 
VM, even though according to the specification \emph{m3.medium} has 3 ECUs and 
\emph{m1.small} has 1. Therefore, the previous extrapolation of \emph{m3.medium}'s 
capacity has been an overestimation. Hence, when a new VM is needed again, the algorithm 
selects \emph{m1.small} again.

\begin{figure}[!t]
	\centering
	
	\includegraphics[width=0.8 \linewidth]{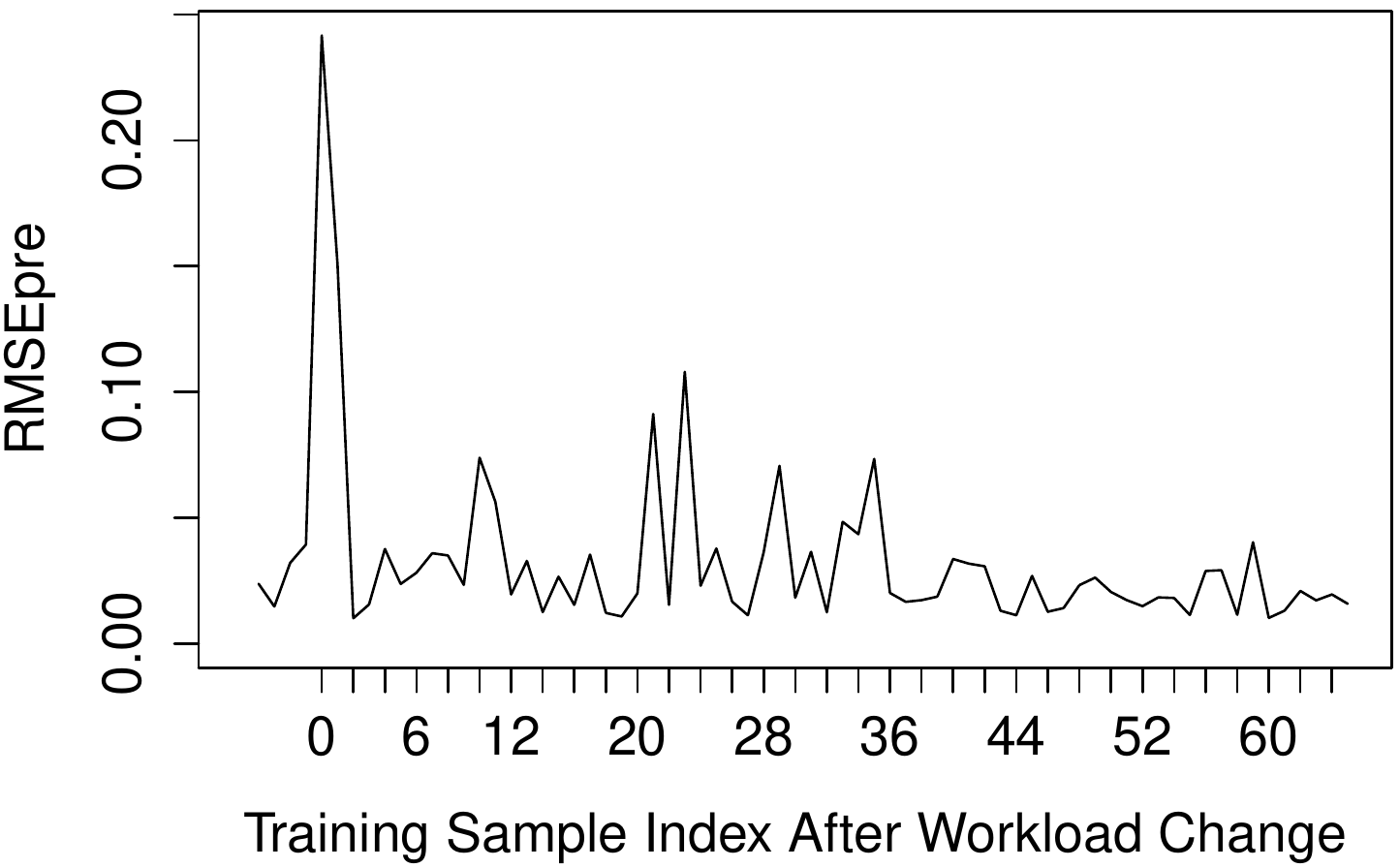} 
	\caption{RMSE-pre in the presence of a workload change. The 0 index corresponds to 
		the first sample after the workload change. }
	\label{fig:RMSE-Inj}
	
\end{figure}

3.5 hours after the start of the experiment the workload change is injected. This is 
reflected in the HTMs' anomaly scores $an_{k}$ and the ANN's errors. Consequently, the 
\emph{learning rate} $lr_{k}$, the \emph{momentum} $m_{k}$ and the \emph{epochs} $e_{k}$ 
also change to speed up the learning process as per equations~\ref{eq:lrk},~\ref{eq:mk} 
and~\ref{eq:ek} and as a result the ANN adapts quickly to the workload change. As 
discussed for each sample we compute its error (RMSE-pre) before updating the ANN. 
Figure~\ref{fig:RMSE-Inj} depicts how these errors increase when the change is injected 
and decrease afterwards as the ANN adapts timely.

Eventually the load increases enough so the system needs to scale up again. Due to the 
injected change, the workload has become much more memory intensive, which is reflected 
in the ANN's prediction. Hence \emph{m1.small} can serve just a few users, given it has 
only 1.7GB RAM. At that point the CPU capacity of \emph{m1.medium} is inferred from the 
capacities of \emph{m1.small} and \emph{m3.medium} as per Eq.~\ref{eq:cpuCapEst}, since 
it has not been used before. Consequently Algorithm~\ref{algo:vmtselect} selects 
\emph{m1.medium} for the 4th VM just before the experiment completes.

\begin{figure}[!t]
	\centering
	
	\includegraphics[width=0.75 \linewidth]{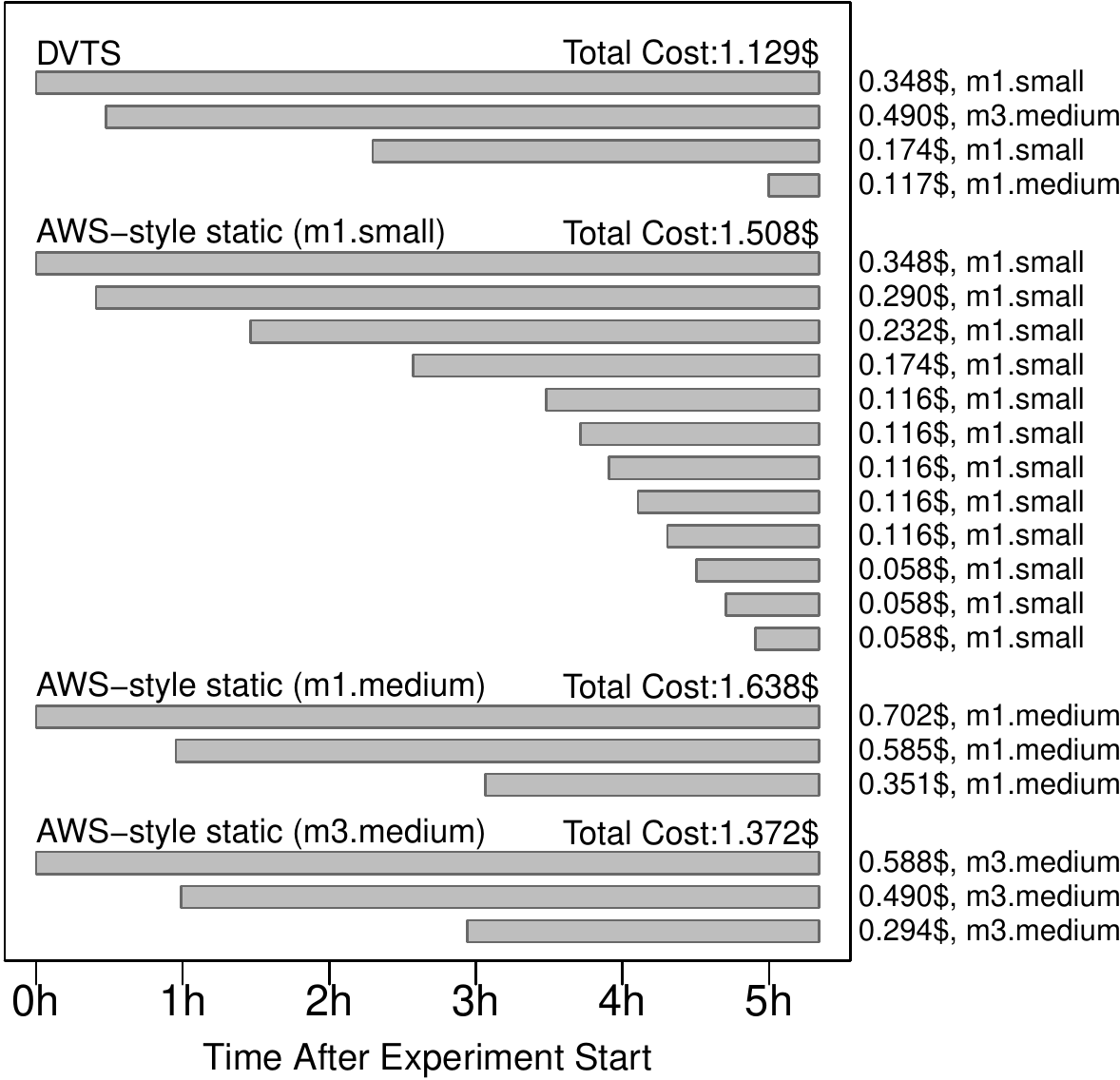} 
	\caption{Timelines and costs of all VMs grouped by experiments. DVTS is our approach. 
	The AWS-style policies are the baselines, which statically select a predefined VM 
	type.}
	\label{fig:Timeline}
	
\end{figure}

For each experiment, Figure~\ref{fig:Timeline} depicts the timelines of the allocated VMs 
and the total experiment costs. For each VM the type and cost are specified to the right. 
Our selection policy is listed as \emph{DVTS}. The baseline policy which statically 
selects \emph{m1.small} allocates 8 new VMs after the workload change as \emph{m1.small} 
can serve just a few users under the new workload. In fact, if there was no \emph{cool 
down} period in the autoscaling, this baseline would have exceeded the AWS limit of 
allowed number of VM instances before the end of the experiment. The baselines which 
select \emph{m1.medium} and \emph{m3.medium} fail to make use of \emph{m1.small} 
instances before the change injection, which offers better performance for money. 

Admittedly, in the beginning DVTS did a misstep with the selection of \emph{m3.medium}, 
because it started with an empty \emph{capacity repository} and had to populate it and 
infer CPU capacities ``on the go''. This could have been avoided by prepopulating the 
\emph{capacity repository} with test or historical data. We could expect that such 
inaccuracies are avoided at later stages, once more capacity and training data is 
present. Still, our approach outperformed all baselines in terms of incurred costs with 
more than 20\% even though its effectiveness was hampered by the lack of contextual data 
in the initial stages.

Our experiments tested DVTS and the baselines with a workload, which is lower than what 
is observed in some applications. While our tests did not allocate more than 12 VMs (in 
the baseline experiment, which statically allocates \emph{m1.small}) many real world 
systems allocate hundreds or even thousands of servers. We  argue that in such cases, 
DVTS will perform better than demonstrated, as there will be much more training data and 
thus the VM types' capacity estimations will be determined more accurately and the 
machine learning approaches will converge faster. As discussed, that would allow some of 
the initial missteps of DVTS to be avoided. Moreover, as the number of AS VMs grows, so 
does the cost inefficiency caused by the wastage of allocated resources, which can be 
reduced by DVTS. 

Finally, the response times in the DVTS experiment and all baseline experiments were 
equivalent. All experiments scale up once the AS VMs' utilisations exceed the predefined 
thresholds, and thus never become overloaded enough to cause response delays. The load 
balancer is equally utilised in all experiments, as it serves the same number of users, 
although it redirects them differently among the AS VMs. Similarly, the DB layer is 
equally utilised, as it always serves all users from all AS VMs.  

\section{Conclusions and future work} \label{conclusion}

In this work we have introduced an approach for VM type selection when autoscaling 
application servers. It uses a combination of heuristics and machine learning approaches 
to ``learn'' the application's performance characteristics and to adapt to workload 
changes in real time. To validate our work, we have developed a prototype, extended the 
CloudStone benchmark and executed experiments in AWS EC2. We have made improvements to 
ensure our machine learning techniques train quickly and are usable in real time. Also we 
have introduced heuristics to approximate VM resource capacities and workload resource 
requirements even if there is no readily usable data, thus making our approach useful 
given only partial knowledge. Results show that our approach can adapt timely to workload 
changes and can decrease the cost compared to typical static selection policies.

Our approach can achieve even greater efficiency, if it periodically replaces the already 
running VMs with more suitable ones in terms of cost and performance, once there is a 
workload change. We will also work on new load balancing policies, which take into 
account the actual VM capacities. Another promising avenue is optimising the scaling down 
mechanisms --- i.e. selecting which VMs to terminate when the load decreases.
Also, we plan to extend our approach, which currently optimises cost, to also consider 
other factors like energy efficiency. This would be important when executing application 
servers in private clouds.  
Finally, we plan to incorporate in our algorithms historical data about VM types' 
resource capacity and workload characteristics.

\section*{Acknowledgments}

We thank Rodrigo Calheiros, Amir Vahid Dastjerdi, Adel Nadjaran Toosi, and Simone Romano 
for their comments on improving this work. We also thank Amazon.com, Inc for their 
support through the AWS in Education Research Grant.\\[0.3cm]



\end{document}